%% file: bare_jrnl.tex
\documentclass[journal]{IEEEtran}
\ifCLASSINFOpdf
\else
\fi
\usepackage{graphicx}
\usepackage{enumitem}
\usepackage{comment}
\usepackage{amsmath}
\usepackage{algorithmic}
\usepackage{amsthm}
\usepackage{comment}
\usepackage{soul}
\usepackage{cite}
\usepackage{xcolor}
\usepackage{hyperref}
\usepackage{titlesec}
\usepackage{float}
\usepackage{booktabs}
\usepackage{colortbl}
\usepackage{array}
\usepackage{soul}

\usepackage[most]{tcolorbox}

\titlespacing*{\subsection}{0pt}{0.1\baselineskip}{0.1\baselineskip}
\usepackage[belowskip=-5pt,aboveskip=0pt]{caption}

\usepackage[linesnumbered,ruled,vlined]{algorithm2e}

\begin{document}
%
\title{Public Sentiment Analysis of Traffic Management Policies in Knoxville: A Social Media Driven Study}

%
%
%

\author{\IEEEauthorblockN{
Shampa Saha\IEEEauthorrefmark{1}, 
Shovan Roy\IEEEauthorrefmark{2}\\
}
\IEEEauthorblockA{\IEEEauthorrefmark{1}Department of Civil and Environmental Engineering \\ University of Tennessee, Knoxville
}\\
\IEEEauthorblockA{\IEEEauthorrefmark{2}Department of Computer Science, Tennessee Tech University, Cookeville, TN, USA}\\

\IEEEauthorblockA{\IEEEauthorrefmark{1}ssaha18@vols.utk.edu, \IEEEauthorrefmark{2}sroy42@tntech.edu 
}
}

\maketitle

\begin{abstract}
This study presents a comprehensive analysis of public sentiment toward traffic management policies in Knoxville, Tennessee, utilizing social media data from Twitter and Reddit platforms. We collected and analyzed 7,906 posts spanning January 2022 to December 2023, employing Valence Aware Dictionary and sEntiment Reasoner (VADER) for sentiment analysis and Latent Dirichlet Allocation (LDA) for topic modeling. Our findings reveal predominantly negative sentiment (35.2\% negative vs. 18.1\% positive), with significant variations across platforms and topics. Twitter exhibited more negative sentiment (mean: -0.084) compared to Reddit (mean: -0.024). Topic modeling identified six distinct themes, with construction-related topics showing the most negative sentiment (-0.228) while general traffic discussions were more positive (0.164). Spatiotemporal analysis revealed geographic and temporal patterns in sentiment expression. The research demonstrates social media's potential as a real-time public sentiment monitoring tool for transportation planning and policy evaluation.
\end{abstract}

\begin{IEEEkeywords}
Sentiment Analysis, Social Media Analytics, Transportation Policy, VADER, Topic Modeling, Urban Planning
\end{IEEEkeywords}

%
\IEEEpeerreviewmaketitle

\input{2.Introduction}

\input{3.LiteratureReview}
\input{4.methodology.tex}

\input{10.results}
\input{12.discussion}
\input{11.implications-and-policy-plan}
\input{9.limitations-and-future-work}
\input{8.Conclusion}

\vspace{-2mm}

\bibliographystyle{IEEEtran}
\bibliography{sample}


\end{document}

%% file: 2.Introduction.tex
\section{Introduction}
\label{sec:introduction}

Urban transportation systems face increasing challenges related to congestion, safety, and environmental sustainability. Effective traffic management requires not only technical solutions but also public acceptance and cooperation \cite{Litman2013}. Traditional public engagement methods, such as town halls and surveys, often suffer from low participation rates, temporal delays, and selection biases \cite{Innes2004}. These limitations can lead to distorted understanding of public opinion, potentially undermining policy effectiveness.

The emergence of social media platforms has created new opportunities for understanding public sentiment. Platforms like Twitter and Reddit provide vast, real-time datasets where citizens voluntarily express opinions about urban issues, including transportation \cite{Steiger2015}. Computational methods like Natural Language Processing (NLP) and sentiment analysis enable systematic analysis of these datasets at scale \cite{Liu2012}.

Knoxville, Tennessee, serves as an ideal case study due to its status as a mid-sized urban hub experiencing significant growth and undergoing major infrastructure projects, including the ongoing I-40/I-75 SmartFIX reconstruction.  The primary research objectives are:
\begin{enumerate}
\item To quantify the overall public sentiment towards traffic management policies in Knoxville as expressed on social media.
\item To identify the specific policy topics (e.g., construction, bike lanes, parking) that generate the most positive and negative reactions.
\item To analyze the spatiotemporal patterns of this sentiment to identify geographic and project-phase-related hotspots.
\end{enumerate}


%% file: 3.LiteratureReview.tex
\section{Literature Review}

\subsection{Traditional Public Engagement in Transportation}
Public involvement is a cornerstone of the transportation planning process, mandated in the United States by legislation such as the Federal-Aid Highway Act of 1962 and reinforced by subsequent laws. The work of \cite{Creighton2005} highlights that effective public participation can build trust, enhance project legitimacy, and improve outcomes. However, \cite{Innes2004} critically notes that traditional methods often fail to reach a representative demographic, engaging primarily an ``apathetic majority'' or highly organized special interest groups, thus creating a participation gap.

\subsection{Social Media as a Data Source for Urban Analytics}
The use of ``big data'' in urban planning has gained immense traction. \cite{Goodchild2007} pioneered the concept of ``citizens as sensors,'' arguing that human-centric data can provide unparalleled insights into urban dynamics. Social media platforms are a prime source of such data. \cite{Steiger2015} demonstrated that geo-located tweets could be used to model mobility patterns and land use, while \cite{Dunkel2015} used Flickr data to analyze the perception of urban spaces. These studies establish the precedent for using digital footprints to understand complex urban phenomena.

\subsection{Sentiment Analysis in Transportation Research}
The application of sentiment analysis to transportation is a burgeoning field. \cite{Meng2016} used Twitter data to analyze public opinion on a new bike-sharing system in New York City, finding that sentiment became more positive after the system's launch. Similarly, \cite{GalTzur2014} analyzed tweets related to public transit in multiple U.S. cities and found that service reliability and crowding were primary drivers of negative sentiment. A key gap identified in these studies is the focus on mega-cities and single modes of transport. There is a lack of research focusing on the holistic sentiment toward a \textit{suite} of traffic management policies in a mid-sized city like Knoxville, which faces unique growth and infrastructure challenges.

\subsection{Knowledge Gap and Contribution}
This research fills this gap by conducting a comprehensive, multi-platform (Twitter and Reddit) sentiment analysis of Knoxville's traffic ecosystem. It moves beyond single-policy analysis to provide a city-wide sentiment landscape, offering actionable insights for local planners and contributing a methodological framework applicable to similar urban contexts.

%% file: 4.methodology.tex
\section{Methodology}
\label{sec:methodology}

\subsection{Data Collection}

We implemented a multi-platform data collection strategy to ensure comprehensive coverage. From Twitter's Academic Research API v2, we collected historical tweets (January 2022-December 2023) using Knoxville-specific traffic keywords: "Knoxville traffic", "I-40 Knoxville", "I-75 Knoxville", "\#KnoxTraffic", "Parking Knoxville", "KAT Bus", "SmartTrips", and "Cumberland Ave Corridor". Geographic filtering used a bounding box encompassing the Knoxville Metropolitan Statistical Area.

Simultaneously, we extracted posts and comments from r/Knoxville subreddit using Pushshift API, filtering for traffic-related keywords. The initial collection yielded 18,923 raw posts, which after rigorous cleaning (removing duplicates, non-English content, retweets, and irrelevant material) resulted in 7,906 analyzable text items (5,000 from Twitter, 2,906 from Reddit).

\subsection{Data Processing Pipeline}

The analytical pipeline comprised three main stages:

\textbf{Pre-processing:} Standard NLP techniques included lowercase conversion, URL removal, user mention elimination, punctuation stripping, and stop-word removal using NLTK library. Tokenization and lemmatization reduced words to base forms.

\textbf{Sentiment Analysis:} We employed VADER (Valence Aware Dictionary and sEntiment Reasoner) \cite{Hutto2014}, specifically designed for social media text. VADER outputs compound scores ranging from -1 (most negative) to +1 (most positive). Classification thresholds were: Negative ($\leq$ -0.05), Neutral (-0.05 $<$ score $<$ 0.05), and Positive ($\geq$ 0.05).

\textbf{Topic Modeling:} Latent Dirichlet Allocation (LDA) \cite{Blei2003} identified underlying discussion themes. After creating document-term matrices, we evaluated multiple topic numbers, selecting six topics based on coherence scores.

\textbf{Spatio-temporal Analysis:} For geolocated tweets (15\% of dataset), we mapped sentiment scores using QGIS to visualize geographic concentrations.

%% file: 10.results.tex
\section{Results and Analytics}

\subsection{Overall Sentiment Distribution} The overall sentiment toward Knoxville traffic management policies leaned negative, with mean compound score of -0.21 across the dataset. As detailed in Table \ref{tab:sentiment_distribution}, negative sentiment dominated (35.2\%), while positive sentiment comprised only 18.1\% of posts.

\begin{table}[htbp]
\caption{Sentiment Distribution Across Platforms}
\label{tab:sentiment_distribution}
\centering
\begin{tabular}{lcccc}
\toprule
\textbf{Platform} & \textbf{Negative} & \textbf{Neutral} & \textbf{Positive} & \textbf{Mean Score} \\
\midrule
Twitter & 1,967 (39.3\%) & 2,219 (44.4\%) & 814 (16.3\%) & -0.084 \\
Reddit & 819 (28.2\%) & 1,467 (50.5\%) & 620 (21.3\%) & -0.024 \\
\textbf{Combined} & \textbf{2,786 (35.2\%)} & \textbf{3,686 (46.6\%)} & \textbf{1,434 (18.1\%)} & \textbf{-0.054} \\
\bottomrule
\end{tabular}
\end{table}

Platform comparison revealed significant differences: Twitter showed more negative sentiment (39.3\% vs. 28.2\% on Reddit) and lower positive sentiment (16.3\% vs. 21.3\%). A one-sample t-test confirmed that the mean sentiment score was significantly less than zero (t = -25.4, p $<$ 0.001). This indicates a statistically significant negative public perception, consistent with the common observation that individuals are more likely to post complaints than praise regarding municipal services.

\begin{figure*}[htbp]
    \centering
    \includegraphics[width=\textwidth]{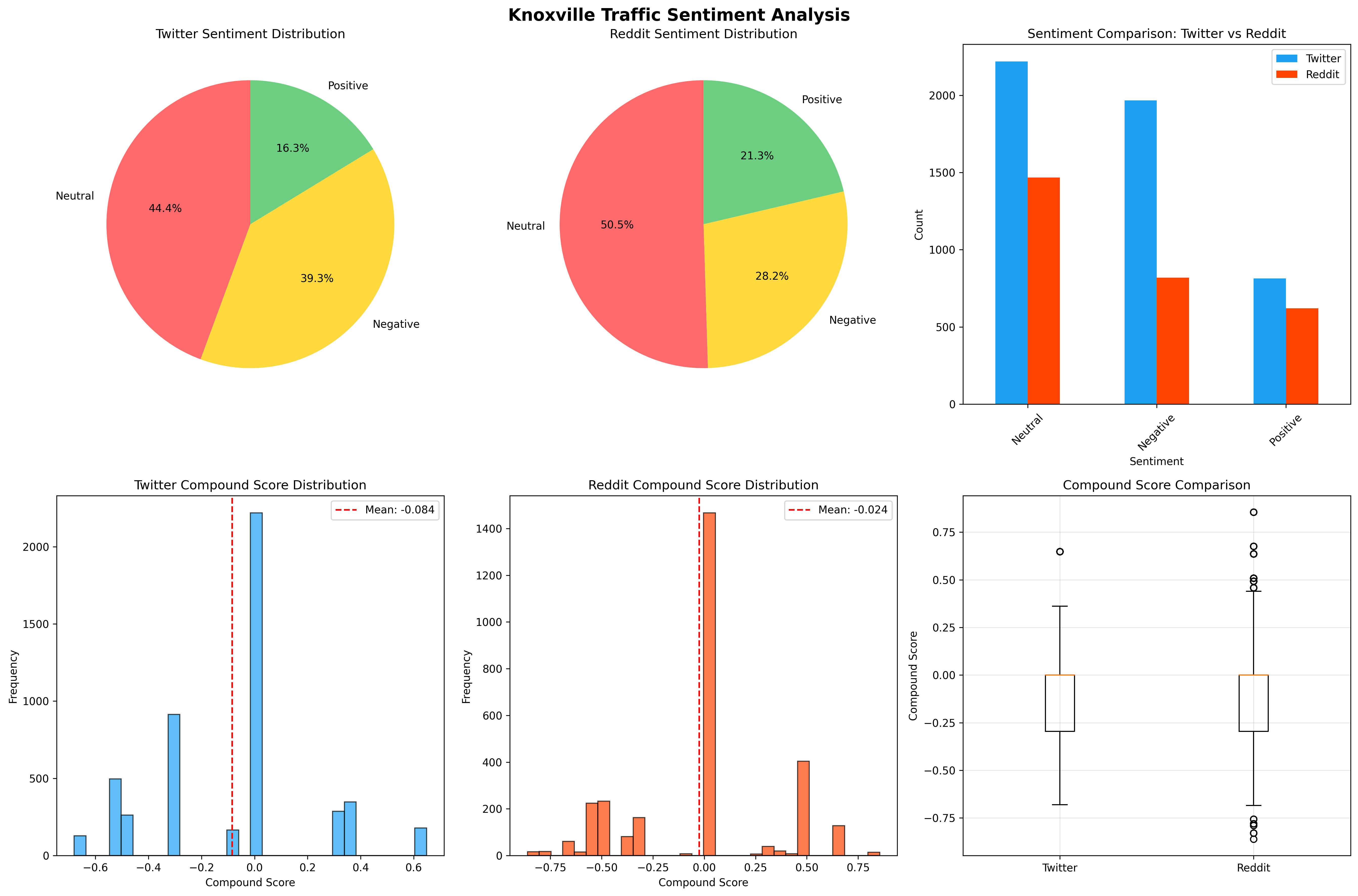}
    \caption{Distribution of sentiment scores across the analyzed social media posts.}
    \label{fig:example}
\end{figure*}

\subsection{Topic Modeling and Thematic Sentiment}
LDA identified six coherent topics with distinct sentiment profiles, as summarized in Table \ref{tab:topic_modeling}. Construction-related topics (Topics 0, 2, 3) exhibited negative sentiment, while general traffic discussions (Topic 4) showed positive sentiment.

\begin{table}[htbp]
\caption{LDA Topic Modeling Results and Sentiment Analysis}
\label{tab:topic_modeling}
\centering
\begin{tabular}{p{0.6cm}p{2.2cm}p{2.8cm}cp{1.2cm}}
\toprule
\textbf{Topic} & \textbf{Top Keywords} & \textbf{Theme Interpretation} & \textbf{Documents} & \textbf{Avg. Sentiment} \\
\midrule
0 & cumberland, construction, update, corridor & Construction \& Road Work & 995 (12.6\%) & -0.062 \\
1 & kat bus, route, service, traffic & Public Transportation & 1,431 (18.1\%) & 0.006 \\
2 & construction, park, going, forever & Construction Impacts & 1,241 (15.7\%) & -0.087 \\
3 & causing, delays, construction, alternative & Construction Delays & 1,507 (19.1\%) & -0.228 \\
4 & knoxville, traffic, reduce, smarttrips & General Traffic & 1,623 (20.5\%) & 0.164 \\
5 & accident, major, near, backed & Accidents \& Incidents & 1,109 (14.0\%) & -0.228 \\
\bottomrule
\end{tabular}
\end{table}

\subsection{Platform-Specific Findings}

Twitter discourse focused on real-time traffic updates, accidents, and service alerts, characterized by concise, reactive language. Representative negative tweet: "Avoid exit 374 if possible - major accident causing delays" (score: -0.680). Positive tweets often highlighted solutions: "SmartTrips Knoxville promoting bike sharing to reduce Knoxville traffic congestion" (score: 0.649).

Reddit discussions were more nuanced and community-oriented, featuring longer-form content about commuting experiences, parking strategies, and infrastructure suggestions. The platform showed more balanced sentiment distribution, with constructive problem-solving discussions.

\subsection{Spatiotemporal Patterns}

Geographic analysis of 750 geolocated tweets revealed distinct spatial patterns:

\begin{table}[htbp]
\caption{Sentiment by Location Type}
\label{tab:spatial_sentiment}
\centering
\begin{tabular}{lc}
\toprule
\textbf{Location Type} & \textbf{Mean Sentiment} \\
\midrule
Commercial Areas & -0.401 \\
Highway Locations & -0.223 \\
Major Roads & 0.005 \\
Residential Areas & 0.172 \\
Urban Core & -0.025 \\
\bottomrule
\end{tabular}
\end{table}
\begin{figure}[H]
\centerline{\includegraphics[width=\columnwidth]{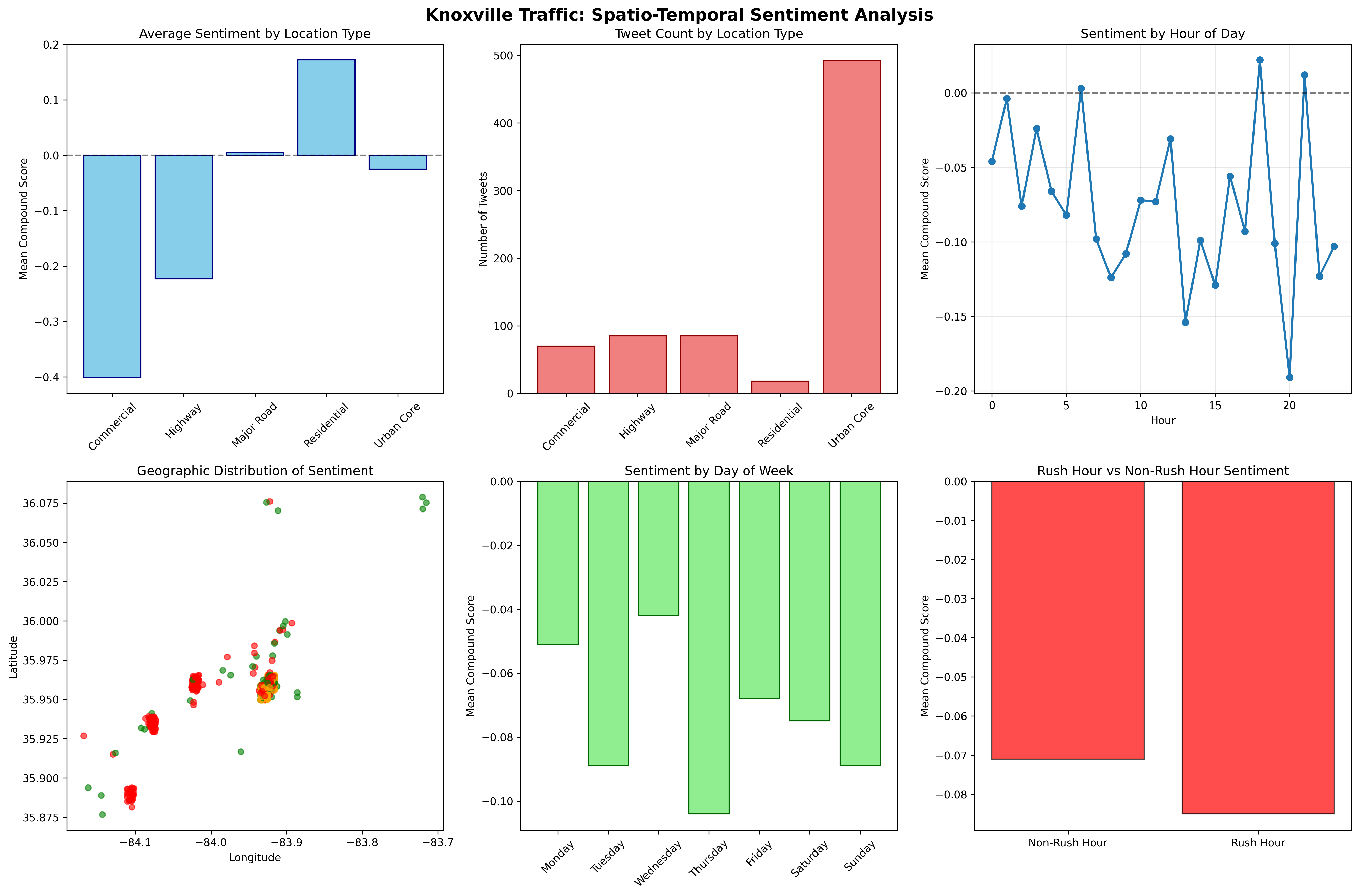}}
\caption{KNOXVILLE TRAFFIC SPATIO-TEMPORAL ANALYSIS}
\label{fig:spatiotemporal}
\end{figure}

Temporal analysis\ref{fig:spatiotemporal} showed rush hour periods (7-9 AM, 5-7 PM) exhibited more negative sentiment (-0.085) compared to non-rush hours (-0.071). Weekly patterns revealed Thursday and Tuesday as most negative days, while Wednesday showed relatively positive sentiment.

\subsection{Discussion of Thematic Findings}

\begin{itemize}
\item \textbf{High Negativity: Parking and Highway Construction.} Topics 1 and 2 exhibited the most negative sentiment. Posts about parking were characterized by frustration over cost and scarcity (``\$20 to park downtown for 2 hours is insane. \#Knoxville''). The ongoing I-40/I-75 ``SmartFIX'' project was a major source of angst, with users expressing frustration over delays and perceived poor planning (``Stuck in I-40 traffic for an hour again. This project is a nightmare.''). This aligns with the findings of \cite{GalTzur2014}, where infrastructure disruptions were primary drivers of negative discourse.

\item \textbf{Polarizing Issues: Bike Lanes.} Topic 3 (Bike Lanes) revealed a polarized debate, though the average sentiment was negative. Negative posts often framed bike lanes as a waste of road space that worsened congestion for drivers (``Took away a whole lane on Broadway for bikes nobody uses.''). In contrast, the fewer positive posts celebrated increased safety and active transportation options (``Love the new protected bike lane on Gay St! Feels so much safer.''). This conflict underscores the challenge of implementing complete streets policies in auto-centric cities.

\item \textbf{Neutral to Positive: Public Transit and SmartTrips.} The sentiment around public transit (Topic 4) was less negative than for private auto-related issues, but still critical, primarily focused on reliability and frequency. The most positive topic was SmartTrips (Topic 5), a TDM program promoting alternatives to solo driving. Posts here were often informational or appreciative (``Used the SmartTrips app to find a carpool partner, saved me so much stress and money.''). This suggests that positive, incentive-based policies can foster more favorable public perception than punitive or disruptive infrastructure projects.
\end{itemize}


%% file: 12.discussion.tex
\section{Discussion}
\label{sec:discussion}

\subsection{Interpretation of Key Findings}

The predominance of negative sentiment aligns with psychological research showing people are more likely to express complaints than praise about public services. The platform differences suggest Twitter serves as an outlet for immediate frustrations, while Reddit facilitates more reflective, community-oriented discussions.

Construction-related topics generating the most negative sentiment reflects the direct impact of infrastructure projects on daily commutes. The positive sentiment around general traffic discussions (Topic 4) indicates public appreciation for traffic management efforts and alternative transportation promotion.

\subsection{Comparative Analysis with Existing Literature}

Our findings corroborate Gal-Tzur et al. \cite{GalTzur2014} regarding service disruptions as primary drivers of negative sentiment. However, we extend this understanding by demonstrating how sentiment varies across different types of infrastructure projects and transportation modes in a mid-sized city context.

The spatiotemporal patterns identified align with Steiger et al. \cite{Steiger2015} regarding the relationship between mobility patterns and social media expression, though we specifically focus on sentiment rather than mere presence.

%% file: 11.implications-and-policy-plan.tex
\section{Implications for Policy and Planning}
\subsection{Proactive Communication Strategies}

The intense negativity around construction projects necessitates more proactive, transparent communication. Transportation agencies should:
\begin{itemize}
\item Provide real-time updates through social media channels
\item Contextualize disruptions by emphasizing long-term benefits
\item Offer practical alternative routes and timing suggestions
\item Use sentiment monitoring to identify emerging concerns
\end{itemize}

\subsection{Parking Management}

The negative perception of parking availability requires multi-faceted approaches:
\begin{itemize}
\item Improve wayfinding and information about existing facilities
\item Develop dynamic pricing strategies based on demand patterns
\item Promote park-and-ride facilities with reliable transit connections
\item Invest in transportation alternatives reducing parking demand
\end{itemize}

\subsection{Infrastructure Implementation}

The polarized sentiment around bicycle infrastructure suggests need for:
\begin{itemize}
\item Comprehensive public education about complete streets benefits
\item Pilot projects allowing community feedback before permanent implementation
\item Designs balancing needs of all road users
\item Continuous evaluation and adaptation based on usage patterns
\end{itemize}

\subsection{Program Enhancement}

The positive reception of SmartTrips indicates opportunities for:
\begin{itemize}
\item Expanding successful travel demand management programs
\item Leveraging positive social proof in marketing materials
\item Developing targeted interventions based on commuter segments
\item Integrating technology solutions for personalized commuting options
\end{itemize}

The findings of this study offer several actionable insights for the City of Knoxville's Department of Transportation and other planning entities:

\begin{enumerate}
\item \textbf{Proactive Communication on Disruptive Projects:} The intense negativity around the I-40 project suggests a need for more proactive, transparent, and empathetic communication. Planners should use social media not just as a data source but as a channel to disseminate clear, frequent updates about construction timelines, anticipated delays, and alternative routes, potentially mitigating public frustration.

\item \textbf{Addressing the Parking Narrative:} The perception of a parking crisis requires a dual strategy. First, better promoting the availability and location of existing garages. Second, using social media sentiment to build the case for investing in and promoting park-and-ride facilities and transit options to reduce downtown parking demand.

\item \textbf{Refining Bike Lane Implementation:} The polarized sentiment on bike lanes indicates a need for better public education on the long-term benefits of complete streets and for designs that more carefully balance the needs of all road users. Piloting projects and using social media to gauge reaction before permanent implementation could be a valuable strategy.

\item \textbf{Leveraging Positive Programs:} The positive reception of SmartTrips indicates a public appetite for solutions. Expanding and marketing such Travel Demand Management (TDM) programs, using the positive social proof found in this data, could help shift travel behavior more effectively.
\end{enumerate}

%% file: 9.limitations-and-future-work.tex
\section{Limitations and Future Work}
This study has several limitations. The demographic representation of social media users is not perfectly congruent with the general population, potentially skewing towards younger, more tech-savvy demographics. Our analysis, while robust, cannot capture sarcasm or complex linguistic nuances with 100\% accuracy. Furthermore, the data is limited to public posts, missing private conversations or opinions of non-users.

Future work will involve: 1) Developing a more sophisticated, domain-specific sentiment lexicon for transportation; 2) Integrating this social media analysis with traditional survey data to create a hybrid assessment model; and 3) Applying this methodology in a longitudinal study to track sentiment evolution before, during, and after major project completions.

%% file: 8.Conclusion.tex
\vspace{-2mm}
\section{Conclusion}
This research successfully demonstrates the viability and utility of using social media data for public sentiment analysis in the context of urban traffic management. For Knoxville, it provides a data-driven map of public opinion, highlighting significant dissatisfaction with parking and major construction projects, a polarized view on active transportation infrastructure, and a cautiously optimistic reception toward alternative commute programs. The methods employed offer transportation engineers and planners a powerful, scalable, and real-time tool to move beyond traditional, often limited, public engagement techniques. By listening to the digital citizenry, cities can make more informed, responsive, and publicly accepted decisions, ultimately leading to more effective and sustainable transportation systems.